# High-resolution single-shot spiral diffusion-weighted imaging at 7T using expanded encoding with compressed sensing

Gabriel Varela-Mattatall*[1,2], Paul I. Dubovan*[1,2], Tales Santini[1,2], Kyle M. Gilbert[1,2], Ravi S. Menon[1,2], Corey A. Baron[1,2]

[1] Centre for Functional and Metabolic Mapping (CFMM), Robarts Research Institute, Western University, London, Canada
[2] Department of Medical Biophysics, Schulich School of Medicine and Dentistry, Western University, London, Canada
* Equal contribution

Magnetic Resonance in Medicine (Technical Note: 2800 words + 5 figures/tables)

Total: 2649 > 2800w

# Figures and Tables: 5/5
# Supporting information documents: 5

Submitted to Magnetic Resonance in Medicine

**Abstract**

**Purpose:** The expanded encoding model incorporates spatially- and time-varying field perturbations for correction during reconstruction. So far, these reconstructions have used the conjugate gradient method with early stopping used as implicit regularization. However, this approach is likely suboptimal for low-SNR cases like diffusion or high-resolution MRI. Here, we investigate the extent that l1-wavelet regularization, or equivalently compressed sensing (CS), combined with expanded encoding improves trade-offs between spatial resolution, readout time and SNR for single-shot spiral diffusion-weighted imaging at 7T. The reconstructions were performed using our open-source GPU-enabled reconstruction toolbox, "MatMRI", that allows inclusion of the different components of the expanded encoding model, with or without CS.
**Methods:** *In vivo* accelerated single-shot spirals were acquired with five acceleration factors (2 - 6) and three in-plane spatial resolutions (1.5, 1.3, and 1.1 mm). From the *in vivo* reconstructions, we estimated diffusion tensors and computed fractional anisotropy maps. Then, simulations were used to quantitatively investigate and validate the impact of CS-based regularization on image quality when compared to a known ground truth.
**Results:** *In vivo* reconstructions revealed improved image quality with retainment of small features when CS was used. Simulations showed that the joint use of the expanded encoding model and CS improves accuracy of image reconstructions (reduced mean-squared error) over the range of acceleration factors investigated.
**Conclusion:** The expanded encoding model and CS regularization are complementary tools for single-shot spiral diffusion MRI, which enables both higher spatial resolutions and higher acceleration factors.

## 1 | Introduction

In an ideal setting, the Fourier transform effectively maps image space into k-space. The Fourier transform provides optimal conditioning of the reconstruction problem, and fast image reconstruction by its inverse Fourier transform.[1–3] However, in practice, artifacts occur when unwanted field perturbations that arise from eddy currents and field inhomogeneity/drift are present. Long readout times and aggressive hardware usage, such as the use of diffusion gradients, exacerbate these issues due to increased phase errors and stronger eddy currents, respectively, which leads to artifacts like ghosting, blurring and geometric distortion.[1–5] Notably, non-Cartesian k-space trajectories[2,3,6,7] and ultra-high-field imaging[5,6] are more susceptible to artifacts from non-ideal fields. Although there are several retrospective image-based methods to compensate for specific kinds of artifacts (for example, diffusion gradient eddy-current correction[8]), these solutions typically apply to Cartesian trajectories.

The expanded encoding model[1–5] incorporates static and dynamic field evolution to correct for unwanted field perturbations during the reconstruction process. A $B_0$ map allows for the compensation of static field offsets, whereas dynamic fields can originate from gradient coils, eddy currents, and physiological events. By monitoring a sequence in real-time with NMR field probes, the field dynamics can be characterized using spherical harmonics up to 2$^{nd}$ or 3$^{rd}$ order in space. Field monitoring has been shown to improve image quality for Cartesian EPI in diffusion[9] and functional acquisitions[10] at 7T, as well as with spiral acquisitions. [1–6]

To date, the expanded encoding model has been incorporated in reconstructions via least-squares (LS) optimizations with no explicit regularization. Instead, reconstructions from previous works have utilized early stopping of conjugate gradient iterations as a form of implicit regularization.[5] However, choosing the correct number of iterations requires manual investigation of image quality at different iterations, and in lower SNR sequences such as diffusion MRI, noise amplification from improper selection of the number of iterations can have severe effects on the quality of computed parameter maps. Furthermore, least-squares optimization is unlikely to accurately solve low SNR problems if not accompanied by a regularization term. Therefore, in this work we investigate complementing the expanded encoding model, $A_{\text{exp}}$, with $\ell_1$ regularization (i.e., compressed sensing (CS))[11–13] to make reconstructions more robust to noise and to eliminate the need for manual tuning of the number of iterations:

$$\hat{x} = \text{argmin}_x \quad \|A_{\text{exp}}x - y\|_2^2 + \lambda\|Wx\|_1, \quad (1)$$

where $x$ is the image, $y$ is the acquired k-space data from the scanner, $\lambda\|Wx\|_1$ is the regularization term that promotes sparsity in the reconstructed image, $\hat{x}$, using wavelets, W, as the sparsifying transform, and $\lambda$ is the regularization weighting. Although the determination of $\lambda$ is not trivial, several approaches have been proposed for its automatic selection[14–16]. CS reconstructions with expanded encoding were implemented using the MatMRI package, which is the only public toolbox we know of for expanded encoding model reconstructions, with or without CS. Through *in vivo* data and simulations, we show that these regularized reconstructions improve image quality compared to early stopping for single-shot spiral diffusion MRI at 7T.

2 | Methods

2.1 Reconstruction

All reconstructions were performed using our open-source MatMRI toolbox, where "Mat" signifies extensive usage of matrix-vector operations in the reconstruction process (https://doi.org/10.5281/zenodo.4495476). MatMRI is a GPU-enabled reconstruction framework, programmed in Matlab (MathWorks), that can handle any k-space trajectory with the option to include a $B_0$ map, higher-order coefficients fitted from the monitored field dynamics, and $\ell_1$-norm regularization. Matlab was chosen as the coding environment due to its prevalence in the MRI community and its straightforward GPU implementation that requires no other libraries or compilation steps. MatMRI only uses standard Matlab libraries and toolboxes to eliminate the need for compilation of binary files, which simplifies installation and usage.

We implemented $\ell_1$-norm regularization using the undecimated Wavelet transform[17,18], W, and the determination of the regularization weighting, $\lambda$, is automatically determined from $WA^T y$, similar to our earlier work.[14] Supporting information figure S1 shows a comparison between different methods to determine $\lambda$ that were evaluated with different numbers of wavelet levels and spatial resolutions from *in vivo* data reconstructions. From these preliminary comparisons, we decided to perform experiments with one wavelet level and to define $\lambda = \sigma/2$, where $\sigma$ is the standard deviation of noise plus noise-like artifacts in the high-pass wavelet coefficients (see supporting information figure S1 for more details).

Implicitly regularized LS was implemented using the conjugate gradient method[19] with early stopping of iterations and CS (i.e., wavelet regularized LS) was implemented using balanced FISTA.[17,20] Supporting information figure S2 shows how the number of conjugate gradient iterations impacts the trade-off between artifacts and noise amplification.[6,21] From these preliminary comparisons we chose to run the conjugate gradient method with 20 iterations for all experiments. Supporting information figure S2 also shows an evaluation regarding the number of iterations for balanced FISTA, as it presents a trade-off between reconstruction quality and speed. From this evaluation, we determined 100 iterations provided reasonable reconstruction quality in a permissible reconstruction time.

All experiments were run as 2D reconstruction problems on a workstation with an Intel i9-7900x processor and 128GB RAM with GPU Nvidia GeForce GTX 1080 Ti with 11 GB of GDDR5X memory.

2.2 MRI Acquisitions

2.2.1 In vivo

This study was approved by the Institutional Review Board at Western University and informed consent was obtained prior to scanning. Scanning was performed on a healthy volunteer on a 7-Tesla head-only MRI scanner (Siemens MAGNETOM, Erlangen, Germany), with 80-mT/m gradient strength and a 400-T/m/s maximum slew rate. Concurrent field monitoring was performed using a radiofrequency coil (32-channel receive and 8-channel transmit) with an integrated 16-channel [19]F commercial field-probe system (Skope Clip-on Camera) to obtain the field dynamics, was fit to 2nd order spherical harmonics.[22] The collected k-space data were coil compressed to 20 virtual coils to improve reconstruction speed,[23–26] and noise correlation between receivers was corrected using pre-whitening before reconstructions.[27]

A Cartesian dual-echo gradient-echo acquisition was used to estimate the $B_0$ map (FOV = 240x240 mm$^2$, spatial resolution = 1.5-mm isotropic, $TE_1/TE_2$ = 4.08/5.10 ms). The $B_0$ map was interpolated to match the in-plane resolution of *in vivo* data. From the first echo, we estimated sensitivity coil maps using ESPIRiT.[28]

We acquired fifteen diffusion-tensor imaging (DTI) protocols with spiral trajectories to evaluate reconstruction performance and computation of fractional anisotropy (FA) maps. The scans consisted of all combinations of acquisitions with acceleration factors from 2x to 6x and in-plane spatial resolutions of 1.5, 1.3, and 1.1 mm—all other sequence parameters remained constant: slice thickness = 3mm, number of slices = 10, TE/TR = 33/2500 ms, FOV = 192x192 mm$^2$. The DTI protocol used monopolar pulsed gradient spin-echo encoding using a b-value of 1000 s/mm$^2$ and 6 directions, plus one b = 0 s/mm$^2$ acquisition. The scan time for each DTI protocol was approximately 1 minute. Following image reconstruction, the MRtrix3[29] package was used to estimate the diffusion tensor and obtain FA maps.

2.2.2 Simulations

To simulate reconstructions for comparisons with a known ground truth, the first echo from the Cartesian gradient-echo acquisition was first used as $x$ in the forward model, $y = A_{\exp}x$, to simulate k-space data. The trajectories used in the forward model were acquired during the 1.5 mm in-plane resolution diffusion MRI scans described above. A slice near isocenter was used due to the presence of several high-contrast regions and $B_0$ variations. Complex white noise was added to the image to produce SNR values of [20,10,5] before applying the forward model to the noisy images to obtain the simulated k-space data; these SNR values are representative of the range produced between low and high b-value diffusion encodings. Supporting information figure S3 shows the ground truth, $B_0$ map, virtual receive-coil sensitivities, trajectory, and estimated higher-order coefficients for both 2- and 4-fold accelerated single-shot spirals.

Reconstructions were performed with and without components of the expanded encoding model ($B_0$ map and 2nd order coefficients) to determine their impact on reconstruction quality. The four settings were reconstructed with both LS and CS methods to analyze the impact of adding wavelet regularization. These simulations were also used to investigate how CS regularization and the expanded encoding model act as complementary tools to improve image quality. Finally, we coil compressed physical coils to [8,16,32] virtual coils to analyze the trade-off between coil compression, reconstruction speed, and reconstruction quality.

To quantitatively evaluate reconstruction quality between reconstructions and the ground truth, we used the normalized root-mean-squared error (NRMSE).

3 | Results
3.1 *In vivo*

Figure 1 shows average diffusion-weighted images for the different spatial resolutions and 2x and 4x acceleration rates. Observable blurring occurs due to either low spatial resolution (partial volume effect) or long readout times (T2* decay), or both. Higher acceleration factors and spatial resolutions mitigates blurring, but this comes at the expense of lowering SNR. Despite being an average across images, it is still possible to appreciate a slight noise level from LS reconstructions, which is noticeably reduced at 4x acceleration using CS reconstruction. Supporting information figure S4 shows the average diffusion-weighted images acquired at 6x acceleration, where it is possible to observe the appearance of artifacts that suggest too aggressive of undersampling.

Figure 2 shows FA maps obtained from reconstructions with different acceleration factors and in-plane spatial resolutions. CS-based reconstructions improve SNR for all cases when compared to their LS counterparts and as SNR decreases either by acceleration rate or spatial resolution these differences become more apparent. Despite improvements from CS, at 6x acceleration both methods show an inability to preserve diffusion metric integrity.

Figure 3 shows a comparison between FA maps obtained from spirals with nearly matched readout times. SNR loss when simultaneously accelerating and increasing the spatial resolution is ameliorated using CS, whereby CS reconstruction demonstrates better preservation of FA features than the LS reconstruction at lower SNR cases.

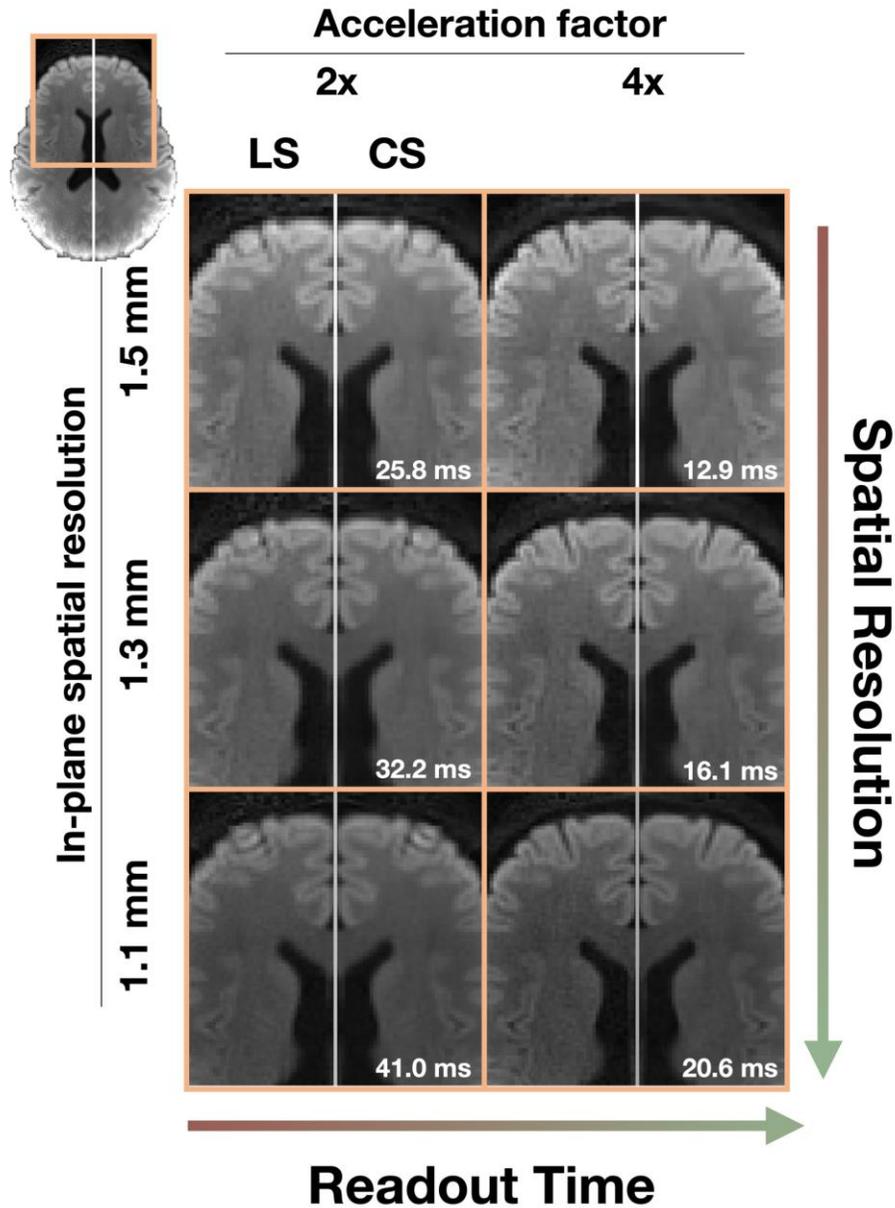

Figure 1- Blurring effect as a function of spatial resolution and readout time. Spatial resolution contributes to blurring due to partial volume effects whereas readout time contributes to blurring due to T2* decay. In each zoomed region, the left side shows the average diffusion-weighted image from least-squares reconstructions (LS), and the right side shows the average diffusion-weighted image from compressed sensing reconstructions (CS). In all panels, the right hemisphere of the averaged image from LS was reflected onto the left side to directly compare anatomy between LS and CS averaged images.

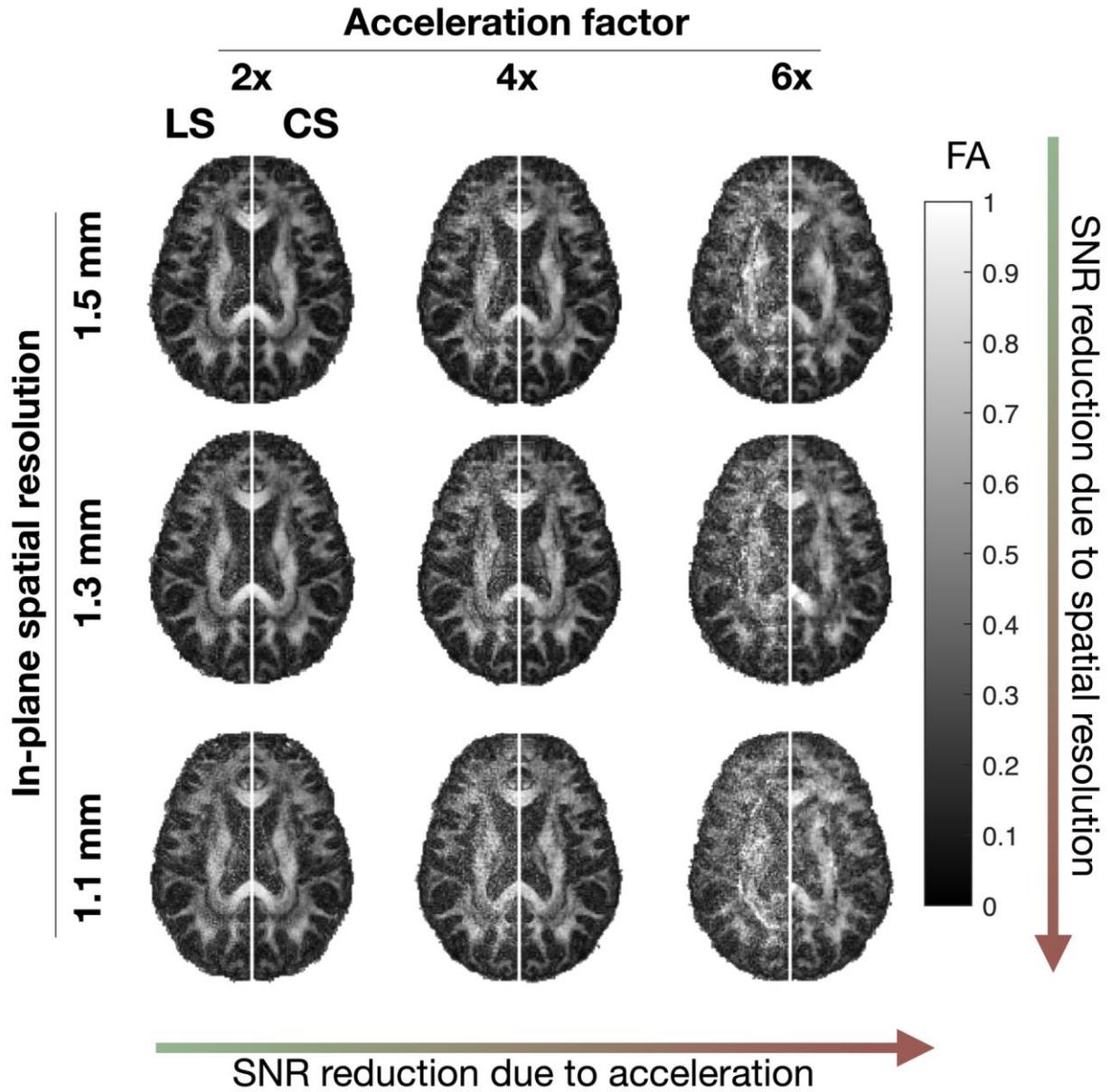

Figure 2- Fractional anisotropy (FA) maps obtained from reconstructions with different acceleration factors and in-plane spatial resolutions. Least-squares (LS) reconstructions were also performed with compressed sensing regularization (CS). In all panels, LS reconstructions from the right hemisphere were reflected onto the left side to directly compare anatomy between LS and CS reconstructions. Top-left reconstruction was performed with the highest SNR condition whereas the bottom-right reconstruction was performed with the lowest SNR condition.

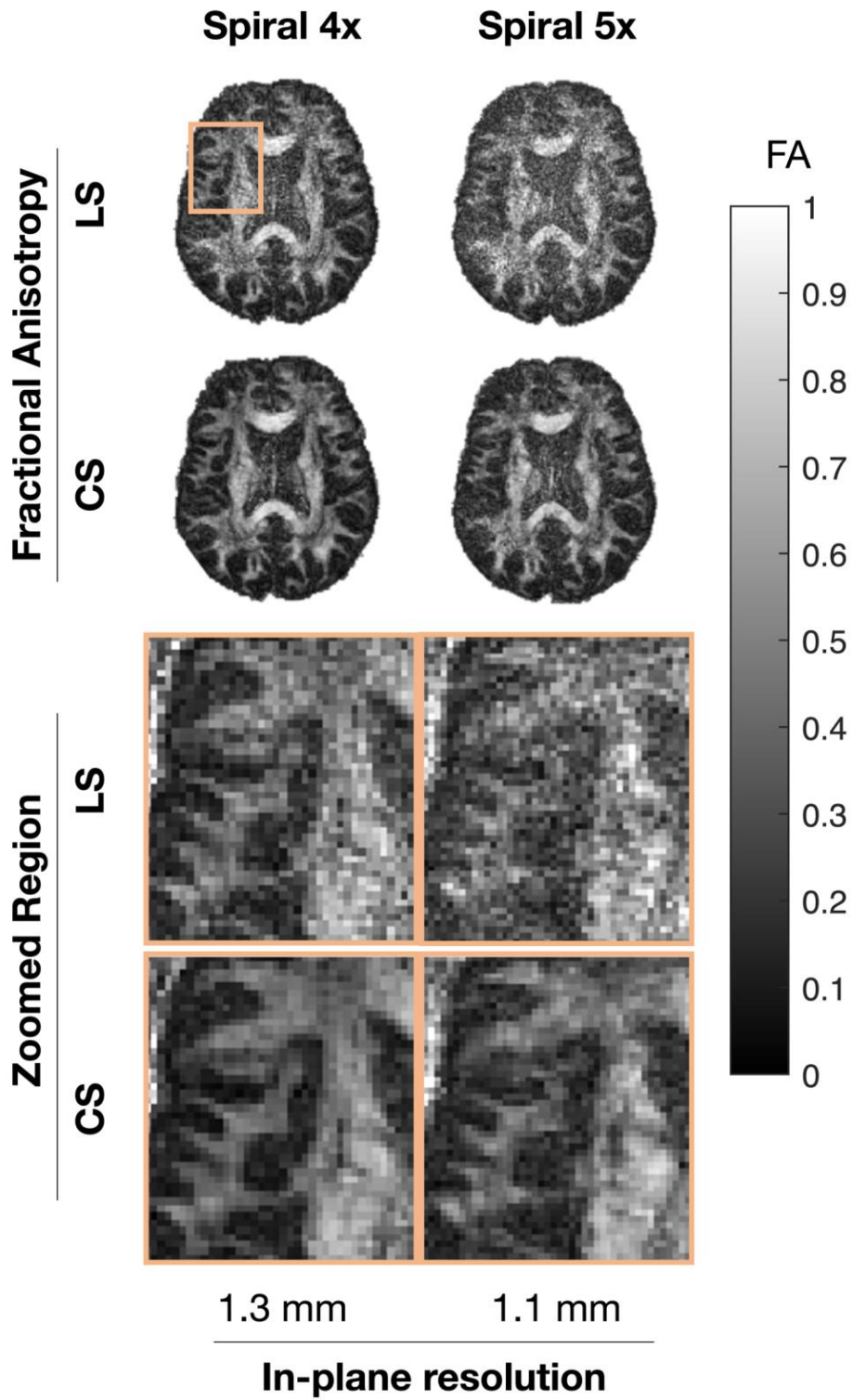

Figure 3- Comparison between spiral protocols based on fractional anisotropy (FA) maps. Their respective combinations of spatial resolution and acceleration rate have nearly matched readout times (16.1 and 16.7 ms, respectively). Image SNR allows for a qualitative comparison, despite images not being co-registered.

3.2 Simulations

Figures 4 and 5 show the summary of the simulations for different combinations of components from the expanded encoding model, acceleration factors, number of virtual coils and noise levels, with and without regularization.

Figure 4 shows the effect of adding $2^{nd}$ order phase terms in the expanded encoding model and CS regularization as a function of the acceleration factor. When the $B_0$ map was not included in the encoding model, the NRMSE was higher than 25%; hence, this case was not included in the figure. At an acceleration factor of 2x, the impact of the expanded encoding model can be appreciated given the noticeably lower NRMSE. The errors from omitting the $2^{nd}$ order terms are reduced when moving to higher accelerations due shorter readout durations (i.e., faster traversal through k-space), reaching a minimum at a factor of 4x. For higher accelerations than 4x, the predominant source of error is noise. At all acceleration factors, the lowest NRMSE occurs for CS with the full expanded model. Supporting information figure S5 shows the reconstructed images for 2x, 4x and 6x acceleration without, with the partial and with the complete expanded encoding model.

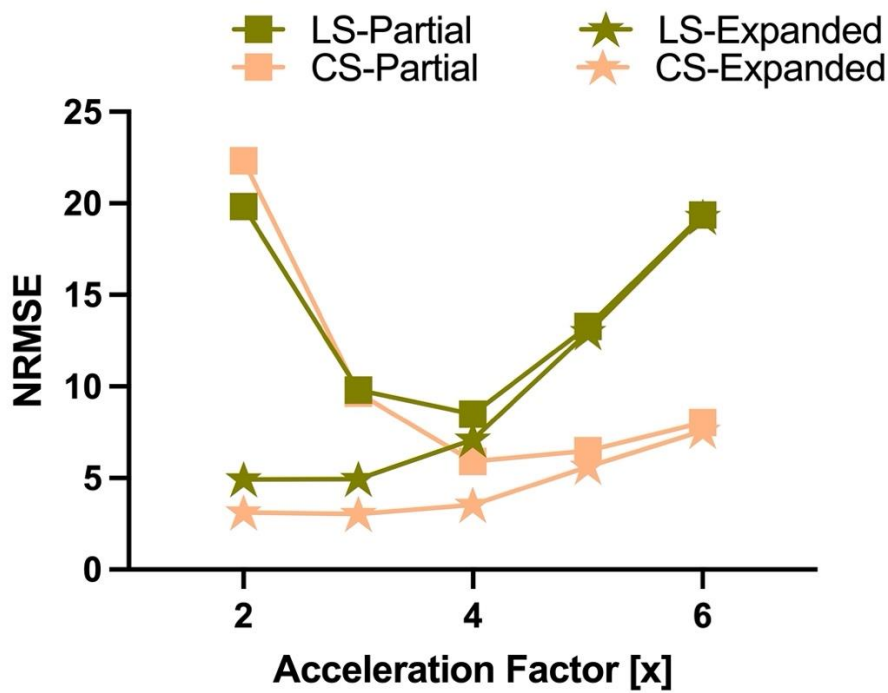

Figure 4- Comparison between least-squares (LS) and compressed sensing (CS) reconstructions. Reconstruction quality, measured in terms of the normalized root-mean-squared error (NRMSE), is presented as a function of the acceleration factor for the case of an SNR of 20 and 16 virtual coils. The "Expanded" case uses the full expanded encoding model, while the "Partial" case does not include the $2^{nd}$ order phase terms.

Figure 5 shows the reconstruction performance for LS and CS reconstructions for the complete expanded encoding model for different numbers of virtual coils. Panel (A) shows NRMSE as a function of the acceleration factor, and panel (B) shows NRMSE as a function of SNR. Both panels illustrate that both LS and CS reconstruction quality are proportional to the number of virtual coils used at either higher acceleration factors or noise levels; however, 16 or more virtual coils only show incremental improvement. When comparing CS to LS, CS reduces NRMSE for all cases with a larger improvement for higher acceleration factors.

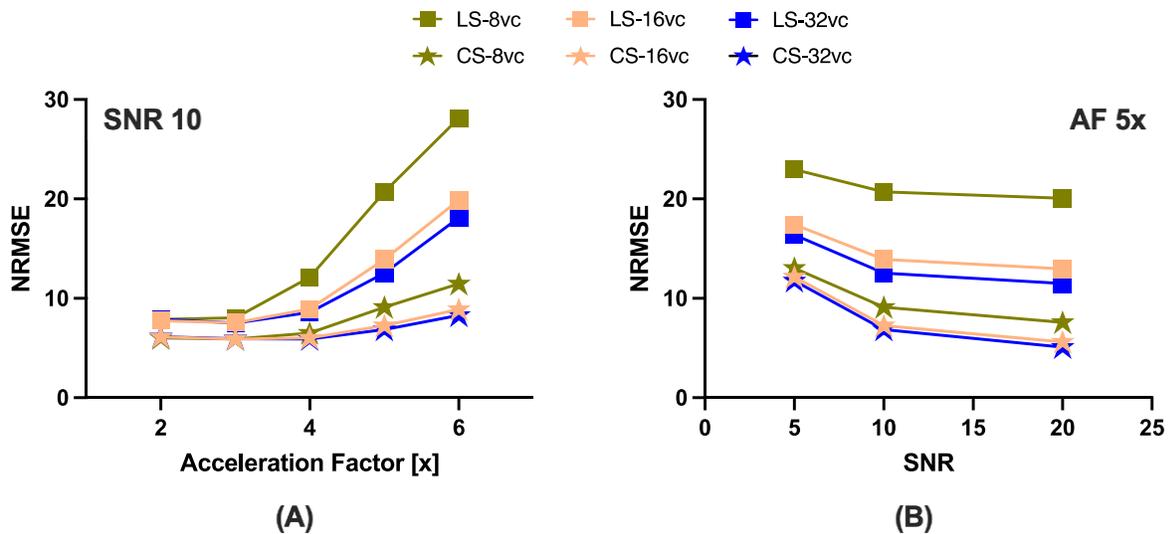

Figure 5- Quality of reconstructions (NRMSE) using the complete expanded encoding model without (LS) and with regularization (CS). Panel (A) shows NRMSE as a function of the acceleration factor whereas panel (B) shows NRMSE as a function of SNR. "vc": virtual coils.

5 | Discussion

In this work, we presented a regularized reconstruction framework that combines the expanded encoding model and CS. The qualitative *in vivo* experiments showed similar trends compared to the quantitative simulations. Together, they demonstrate that the expanded encoding model and CS regularization are complementary tools for improving reconstruction quality. On one hand, at low acceleration factors and high spatial resolution, the main source of error arises from field perturbations given a long readout time. In this acquisition regime, the expanded encoding model provides the greatest benefit during the reconstruction process. However, excessively long readout times may be on the same time order as the lifetime of $^{19}$F field probe signals, which would impair the estimation of higher order coefficients for the expanded encoding model. Additionally, a long readout time also exacerbates T2* blurring that is not accounted for in the expanded encoding model. On the other hand, at high acceleration factors and at high spatial resolutions, the main source of error arises from SNR reduction. In this acquisition regime, CS reconstruction complements the expanded encoding model by using its denoising properties and by removing high-frequency artifacts. In fact, the reduction of aliasing artifacts for 6x acceleration suggests that CS also improves recovery of missing data points for single-shot spirals. When using the automatic selection of regularization weighting factor (see supporting information figure S1), CS did not introduce additional blurring in reconstructed *in vivo* images that can occur with wavelet-based over-regularization. Although our experiments were limited to uniformly accelerated spirals (e.g., artifacts for 6x acceleration), it is expected that variable-density spirals or pseudo-randomized trajectories[30] could improve performance at higher acceleration factors by better transforming aliasing into noise-like artifacts. Furthermore, more sophisticated k-space trajectories are also a motivation for including field monitoring during data acquisition, since they are typically more susceptible to eddy currents (e.g., spiral compared to Cartesian). While Cartesian trajectories like EPI do not have noise-like aliasing artifacts (as opposed to spiral), they would likely still benefit from explicit wavelet regularization that favors solutions satisfying known tendencies of medical images (i.e., sparseness in wavelet domain), instead of implicit early stopping regularization that has poorly defined regularizing behaviour.

In our experiments, coil compression was used to reduce memory requirements and reconstruction time. Results suggest that for the levels of SNR commonly encountered in diffusion MRI, higher acceleration factors should be accompanied by a higher number of virtual coils (≥16) to preserve reconstruction quality, regardless of using LS or CS reconstruction. We recommend using 20-21 virtual coils for a 32-channel receiver as it provides a good trade-off between reconstruction time and image quality, and as shown in related fields[31].

Both LS and CS reconstructions have limitations. For LS reconstruction, the number of iterations must be determined before running the reconstruction. Although the number of iterations according to supporting information figure S2 was quite consistent for both the reconstruction of simulated and *in vivo* data, the non-explicit nature of the regularization is a drawback. Furthermore, the LS reconstruction is strongly limited by SNR reduction that inherently comes in the form of either accelerated acquisitions and high-resolution imaging. For CS reconstruction, the selection of $\lambda$ used here was inspired by our previous work[14] but performed assuming a Rayleigh distribution on the set of wavelet coefficients for one level of the wavelet transform. The additional heuristically determined factor of 1/2 was used to improve reconstruction quality and to avoid blurriness from over-regularization. Despite additional heuristics on the determination of the regularization weighting through a Rayleigh distribution, after its selection the reconstructions showed better performance than LS reconstructions in all cases for different spiral trajectories, acceleration factors, spatial resolutions, virtual coils, and noise levels. Finally, CS requires more iterations than LS with early stopping, which leads to a factor of 5x to 10x increase in reconstruction time (supporting information S2). Future work will consider optimizations to improve reconstruction speed (e.g., unrolled CS with deep learning[32]) and to integrate it in our 7T system for online reconstruction.

6 | Conclusion

In this work we have introduced a compressed sensing extension to the MatMRI toolbox, which was used to investigate the impact of CS on single-shot spiral diffusion MRI at 7T. Simulations and *in vivo* acquisitions showed improved reconstruction quality when combining the expanded encoding model with CS, particularly in low-SNR cases.

ACKNOWLEDGMENT

Authors wish to acknowledge funding from CIHR grant FRN 148453, the NSERC Discovery Grant [RGPIN-2018-05448], Canada Research Chairs [950-231993], Ontario Research Fund [37907], BrainsCAN-the Canada First Research Excellence Fund award to Western University, and the NSERC PGS D program. Finally, authors want to thank Mr. Trevor Szekeres and Mr. Scott Charlton for the data acquisition.

DATA AVAILABILITY STATEMENT

Datasets, experiments, results and MatLab 2019b (MathWorks) codes are available at https://gitlab.com/cfmm/datasets/matmri-4-high-resolution-dmri, and the MatMRI toolbox is publicly available at https://doi.org/10.5281/zenodo.4495476 and https://gitlab.com/cfmm/matlab/matmri.

ORCID

Gabriel Varela Mattatall https://orcid.org/0000-0001-6101-7218
Paul I. Dubovan https://orcid.org/0000-0002-5377-6975
Corey Baron https://orcid.org/0000-0001-7343-5580

TWITTER

Gabriel Varela Mattatall @gabvarelam1
Paul I. Dubovan @pdubovan3
Corey Baron @cabaron1

Supporting information S1: Selection of the Regularization Weighting and Wavelet Level for Compressed Sensing in Spiral Imaging

Comparison between two options for the automatic determination of the regularization weighting (acquisitions are described in Section 2.2). Both methods are based on the principle that the regularization weighting can be determined from the wavelet transform of the zero-filled reconstruction.[14] Both methods assign unique weightings to each wavelet level by considering the histogram of wavelet coefficients in each level (i.e., $\lambda$ is a diagonal matrix), and the weighting for the low-pass filter level is 0.[14]

Method 1 is identical to our earlier work in.[14] In it, it is assumed that the low wavelet coefficient values are dominated by noise and the high wavelet coefficient values are dominated by true tissue signal in the high-pass wavelet levels. The weighting is given by the boundary between these regions, which is determined using k-means clustering with 2 clusters. Zero-valued wavelet coefficients are discarded prior the clustering procedure.

Method 2 also assumes that the low wavelet coefficient values are dominated by noise and the high wavelet coefficient values are dominated by true tissue signal in the high-pass wavelet levels. Here, though, it is assumed that the magnitude of the wavelet coefficient values forms a Rayleigh distribution with a maximum value located at $\sigma$ (which is true when the distributions in each of the real and imaginary channels are Gaussian with a standard deviation of $\sigma$). Upon determining $\sigma$ from the maximum of the histogram of the given wavelet level, the weighting is set to $\lambda = \sigma/2$. While the factor of 2 was heuristically chosen, we found it to provide good image quality for all resolutions and accelerations. Supporting Figure S1a compares reconstruction quality between both methods, and how the selection of the heuristic slightly provides the best reconstruction quality when analyzed with our simulations.

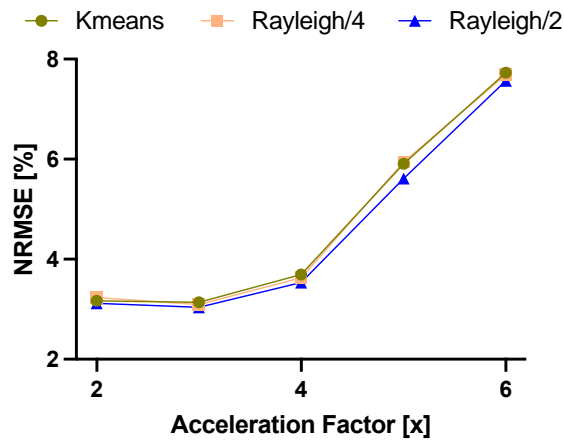

Figure S1a- Effect of the method and its heuristic for image reconstruction at different acceleration factors.

For further investigation, we compared these two options (Kmeans and Rayleigh) with different numbers of wavelet levels (L=1 and L=3) on the *in vivo* data using an undersampling factor of 5x in the 1.3 in-plane spatial resolution case. The L=1 k-means, L=3 k-means, and L=1 Rayleigh methods provided qualitatively similar results, but L=3 Rayleigh exhibited blurring. Finally, we used all DWI reconstructions to compute tensor-based maps such as fractional anisotropy. The slight differences in image quality between the left 3 cases reflects the results from figure S1a. Both quantitative and qualitative results support the selection of L=1 and $\lambda = \sigma/2$ as the method for automatic regularization in spiral imaging.

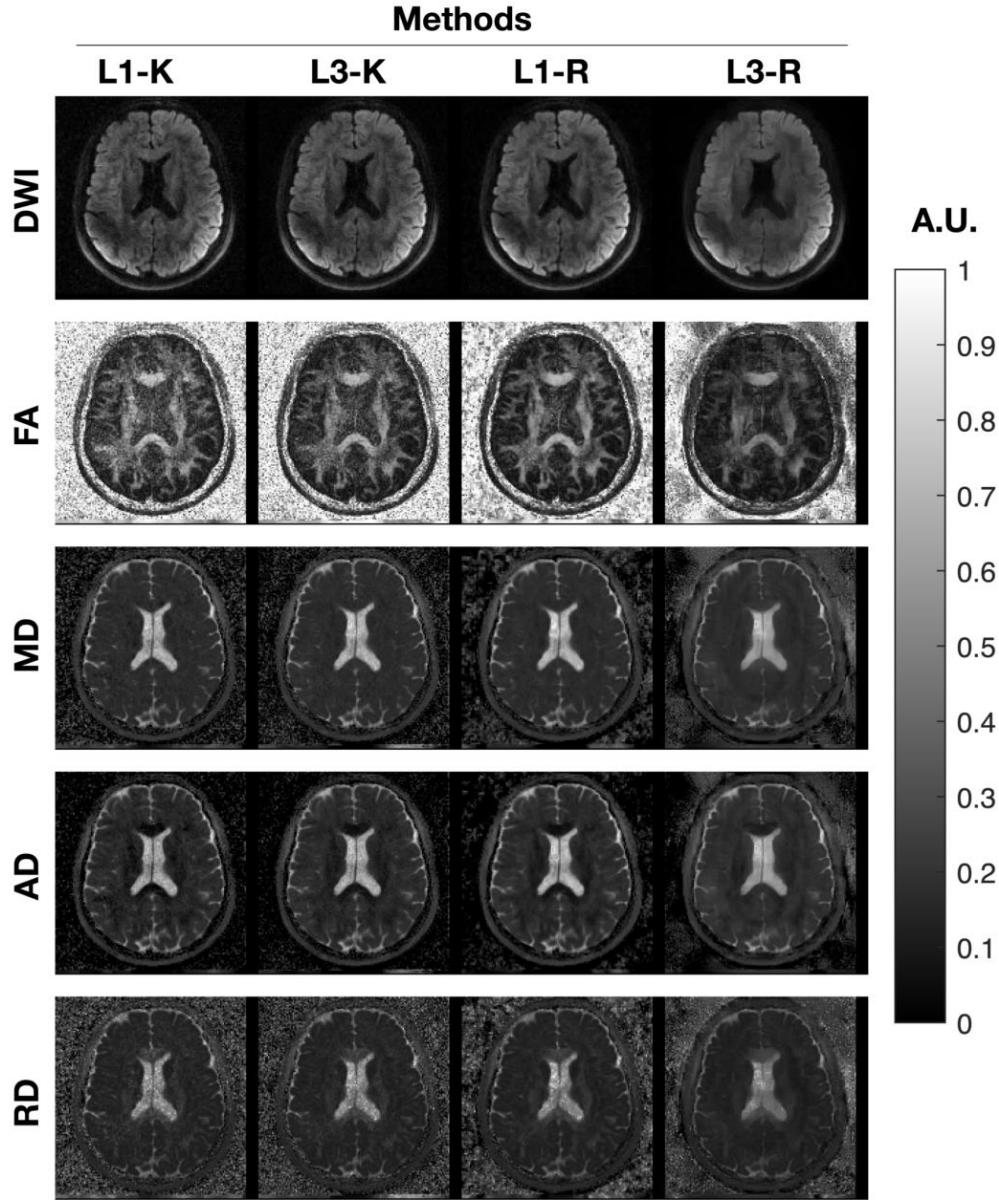

Figure S1b – Effect of the method for the determination of $\lambda$ in compressed sensing reconstructions for spiral imaging. Methods are k-means (K) and $\sigma/2$ (R) with 1 and 3 wavelet levels (L1 and L3, respectively). Panels show diffusion-weighted reconstructions (DWI), and tensor-derived maps like fractional anisotropy (FA), mean diffusivity (MD), axial diffusivity (AD), and radial diffusivity (RD), respectively.

Supporting information 2: Conjugate gradient early stopping criterion and balanced FISTA maximum iteration determination.

To determine the appropriate number of iterations for the conjugate gradient algorithm, we stored the current image reconstruction at [10,15,20,30,40] iterations, and then we visually inspected any noise amplification between image reconstructions. Evaluation was performed in [1.5,1.3,1.1] mm in-plane spatial resolutions and acceleration factors of [2,4,6], since the spatial resolution and the acceleration factor impacts SNR levels, and consequently, it can alter the performance of the algorithm. From analyzing Figures S2a-c, 20 iterations appears to be the upper limit prior noticeable noise amplification, and it provides enough sufficient iterations to ameliorate the aliasing artifact at all acceleration factors.

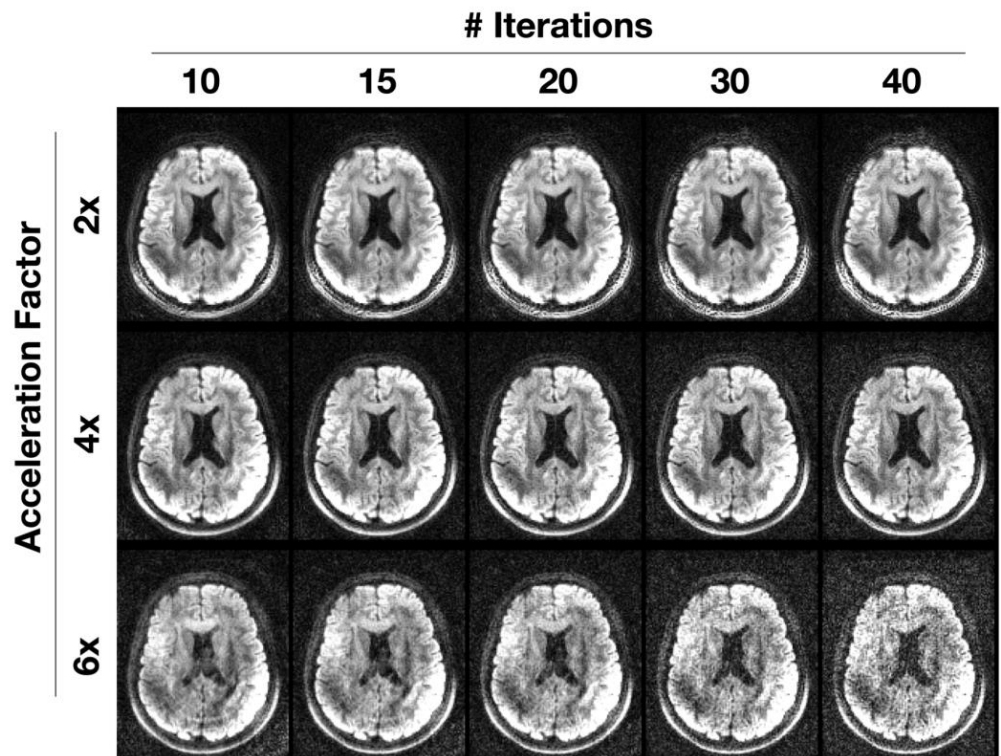

Figure S2a – Evaluation of conjugate gradient algorithm with early stopping criterion based on the number of iterations at 1.5mm in-plane spatial resolution.

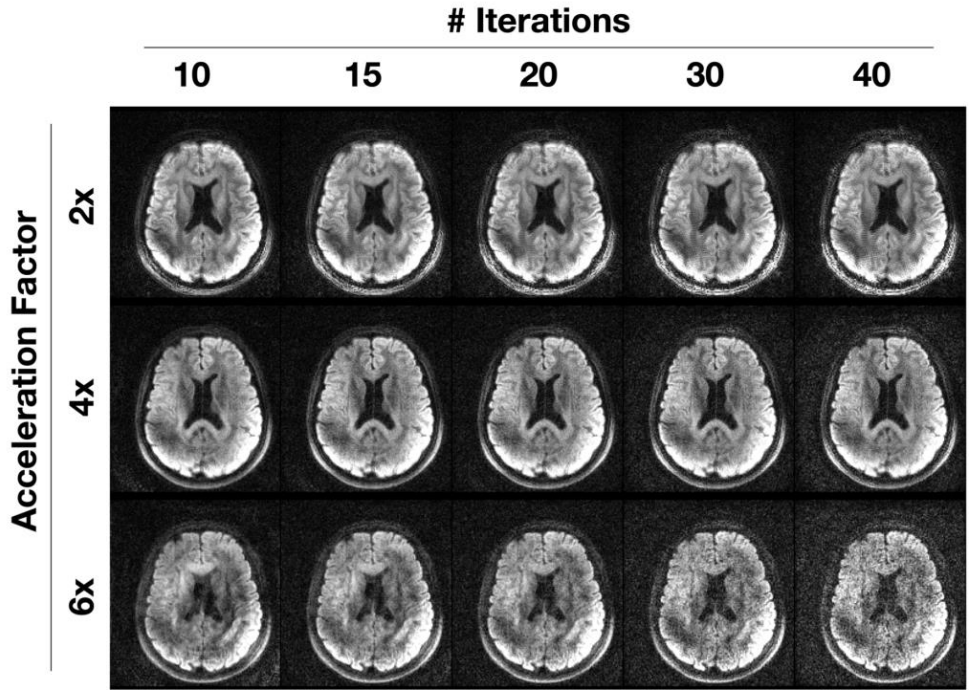

Figure S2b – Evaluation of conjugate gradient algorithm with early stopping criterion based on the number of iterations at 1.3mm in-plane spatial resolution.

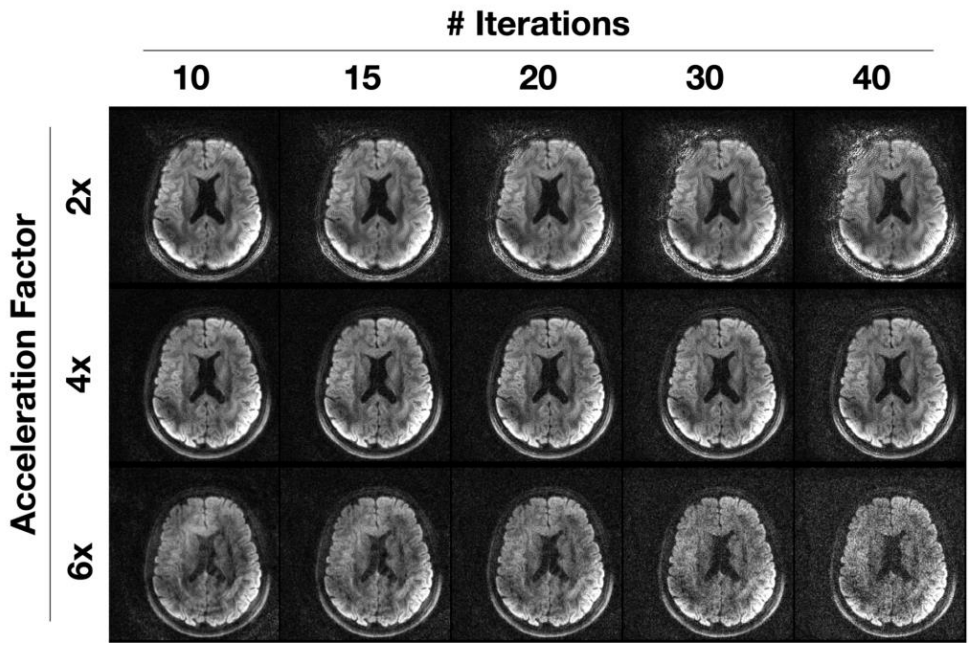

Figure S2c – Evaluation of conjugate gradient algorithm with early stopping criterion based on the number of iterations at 1.1mm in-plane spatial resolution.

To determine the appropriate number of iterations for the balanced FISTA algorithm, images were simulated following the procedure outlined in Section 2.2.2, for [50,75,100,150,200,300] iterations, and for [2,4,6]x acceleration factors. The NRMSE was calculated between the reconstructed images and the ground truth image initially sampled using the forward model. The reconstruction time was also stored for each case. Figure S2d shows that 100 iterations has a good balance between NRMSE and reconstruction time. Figure S2e shows that increasing iterations beyond 100 does not improve the reconstruction at 6x acceleration rate.

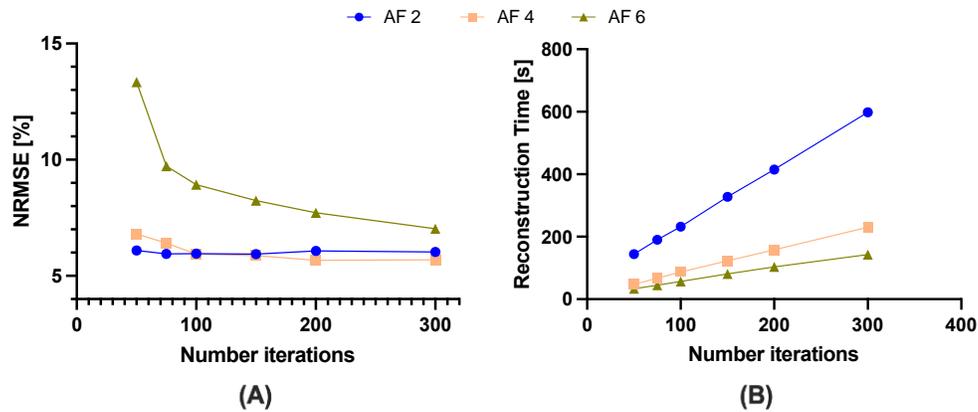

Figure S2d – Performance of reconstructed simulation relative to ground truth image as a function of the number of balanced FISTA iterations. Panel (A) shows reconstruction error whereas panel (B) shows reconstruction time. Reconstructions were based on single-shot spiral diffusion-weighted trajectory with an in-plane resolution of 1.5 mm and acceleration factors (AF) of 2x, 4x and 6x.

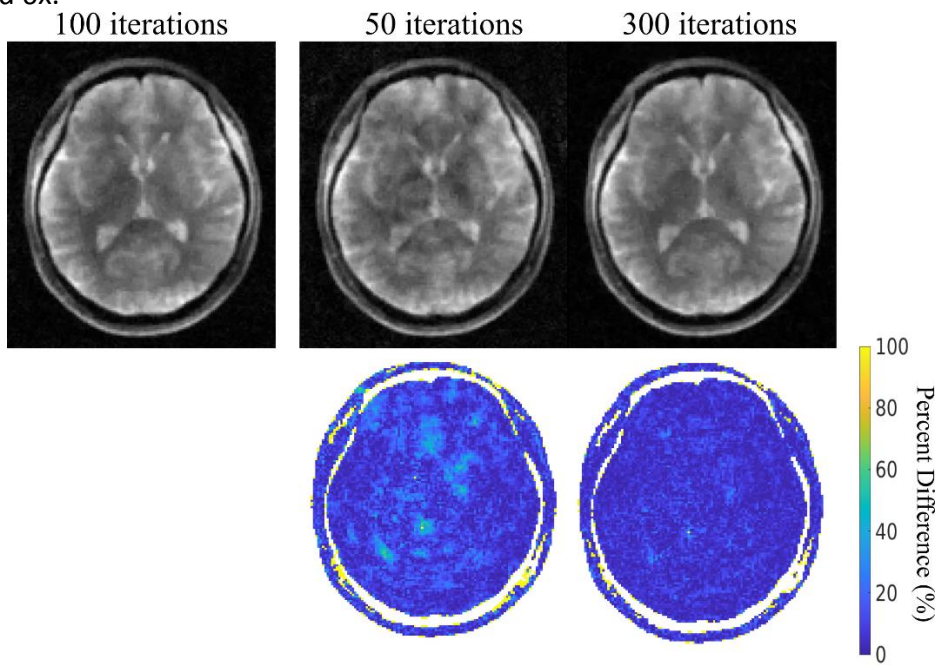

Figure S2e – Reconstructed image simulations using 50, 100, and 300 balanced FISTA iterations for 6x acceleration. Percent difference maps calculated with respect to the image using 100 iterations. No significant improvements in image quality observed beyond 100 iterations.

Supporting Information S3: Expanded encoding model input data for simulations

Supporting information Figure S3 shows the input data for our simulations. Additionally, it also provides the spherical harmonics' coefficients from undersampling factors 2x and 4x to better visualize how they change per acquisition.

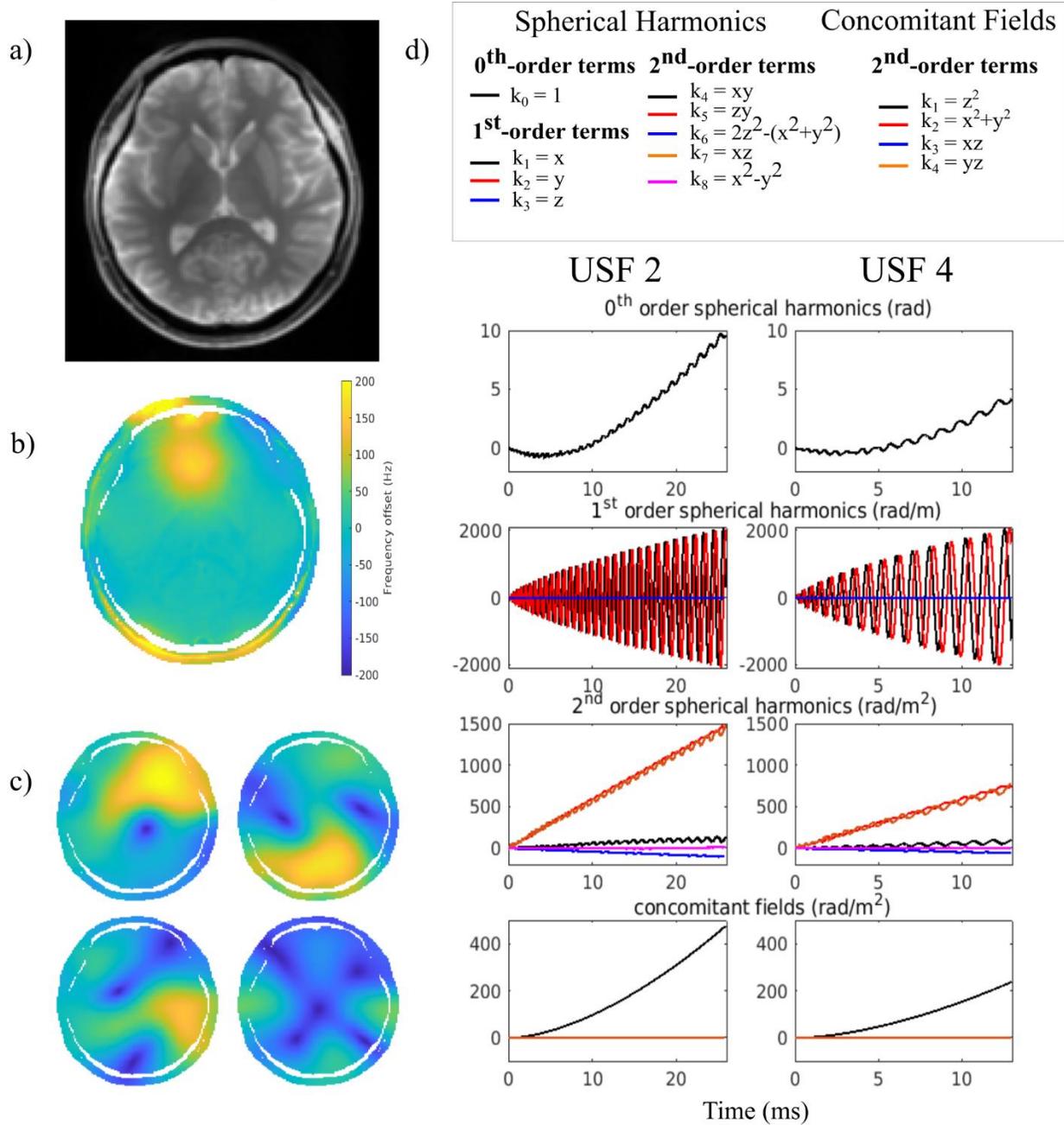

Figure S3: Expanded encoding model input data for simulations. (a) Reference image acquired using a Cartesian dual echo gradient echo sequence. b) Static off resonance field map. c) First four virtual receive coil sensitivities. d) $0^{th}$ to $2^{nd}$ order dynamic spherical harmonic coefficients and second order concomitant fields monitored for diffusion-weighted spiral trajectory readouts acquired at undersampling factors of 2 and 4.

Supporting information 4: Blurring in spiral imaging

In this section we provide complementary information to Figure 1 from the main manuscript. Figure S4 shows the average of the 6 diffusion-weighted acquisitions from 2x, 4x and 6x reconstruction settings performed in our in vivo acquisitions. Both LS and CS reconstructions with an acceleration factor of 6x present artifacts.

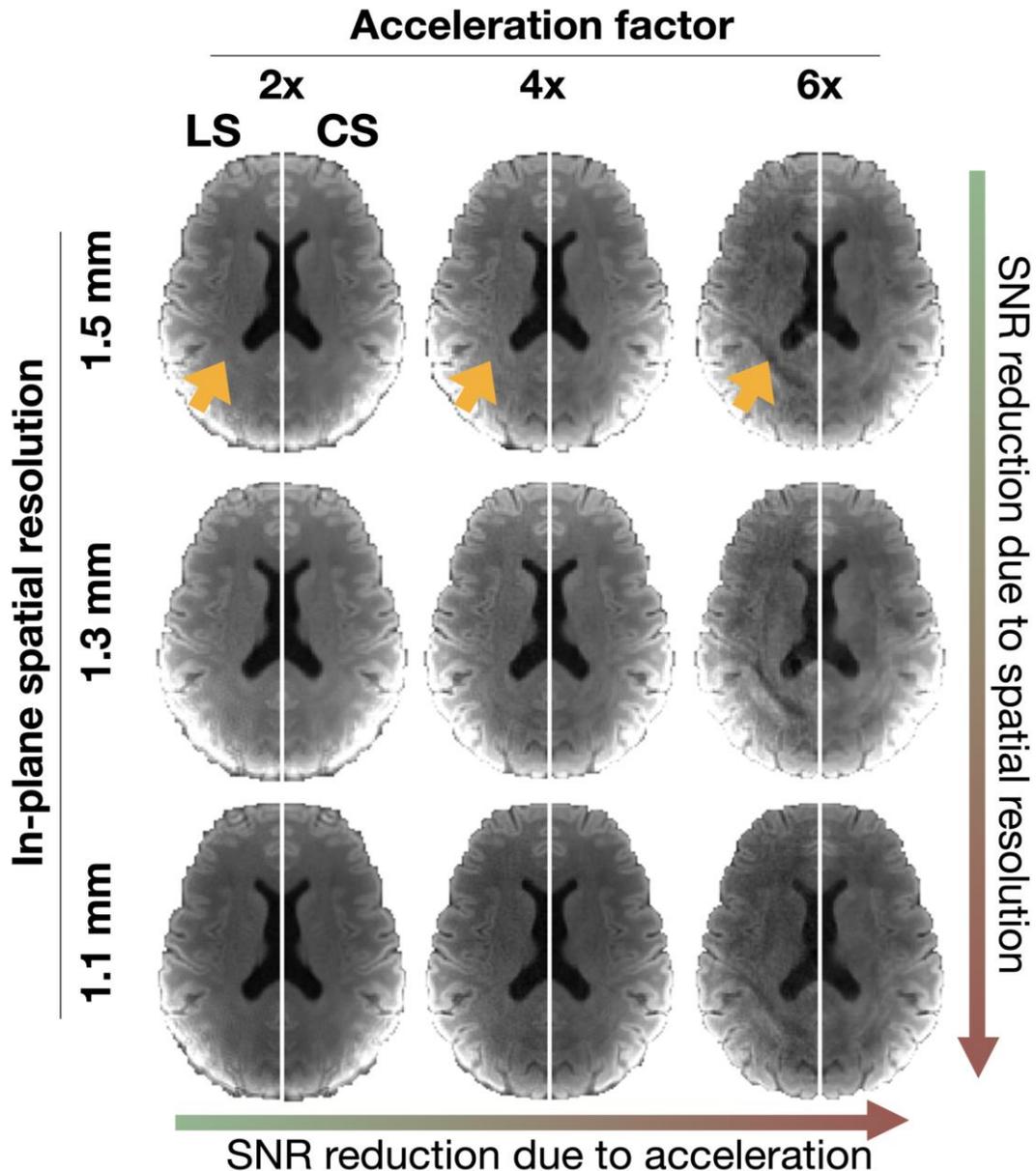

Figure S4- Average of diffusion-weighted images as a function of acceleration and in-plane spatial resolution. The left hemisphere is the reconstruction from least-squares (LS) whereas the right hemisphere is the reconstruction from compressed sensing (CS). The left hemisphere was reflected into the right hemisphere to present the reconstructions from CS. In this set of images, it is clear to observe the SNR reduction due to either acceleration or in-plane spatial resolution. The arrows indicate the location of an artifact observed for the LS reconstruction with an acceleration factor of 6x.

Supporting information figure S5:

Figure S5 shows the effect of both the encoding model and the acceleration factor for LS (left side) and CS (right side) reconstructions using an SNR of 10. With a 2x acceleration rate, the longest readout time generates the largest geometric distortions that are only possible to mitigate by means of the complete expanded encoding model. However, T2* blurring effects remain (peach arrow). With a 4x acceleration rate, the effect from the field perturbations is diminished and by providing a $B_0$ map (second row), most geometric distortions are corrected. However, the lack of $2^{nd}$ order terms results in inaccuracies at the edges farthest from isocenter (olive arrow). Notably, reconstructions do not exhibit noticeable T2* blurring. With a 6x acceleration rate and uniform undersampling, both LS and CS reconstructions fail given the presence of aliasing artifacts, and sets the upper limit in our reconstruction settings. Overall, CS regularization results in improved delineation of anatomical features and higher SNR when combined with the expanded encoding model.

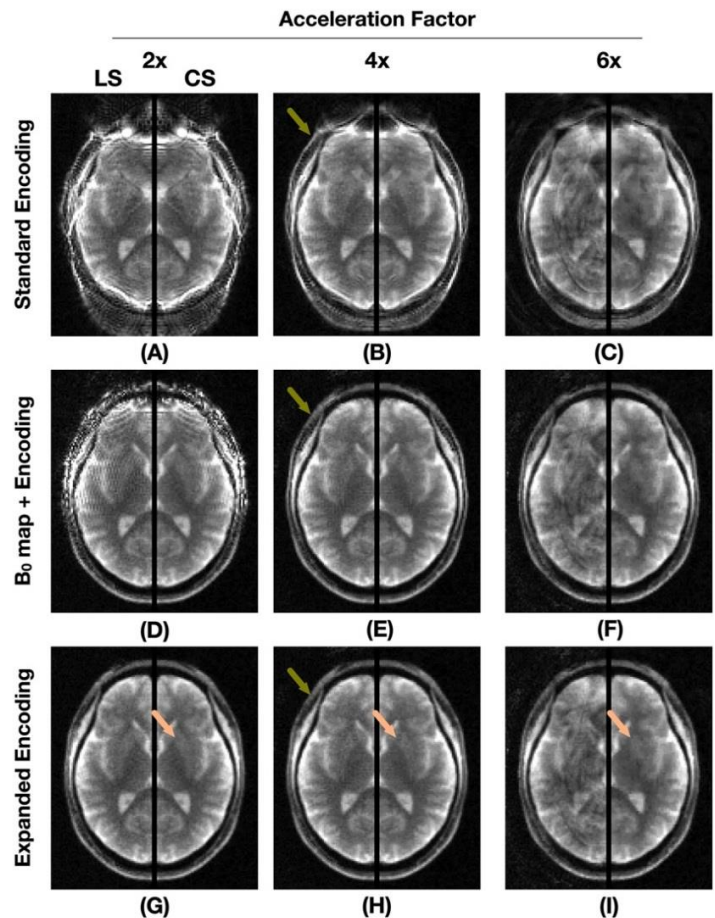

Figure S5- Effect of both the encoding model and the acceleration factor for LS (left side) and CS (right side) reconstructions using an SNR of 10. In all panels, CS reconstructions from the left hemisphere were reflected onto the right side to directly compare anatomy between LS and CS reconstructions. Panels (A-C) are reconstructed with the standard encoding model (i.e., the $B_0$ map and $2^{nd}$-order spherical harmonic terms are not included). Panels (D-F) are reconstructed with the standard encoding model and $B_0$ map, and panels (G-I) are reconstructed with the complete expanded encoding model. Peach arrows depict T2* blurring effects and olive arrows depict the effect from the encoding model in the reconstruction.